\documentclass[preprint]{aastex}

\begin{document}

\title{Emission-line Helium Abundances in Highly Obscured Nebulae}
\author{Mary-Helen Armour}
\affil{Department of Physics and Astronomy, York University, 4700 Keele St., Toronto, ON, Canada~M3J~1P3 \\ Electronic mail: armour@aries.phys.yorku.ca}
\and
\author{David R. Ballantyne\altaffilmark{1}, Gary J. Ferland\altaffilmark{2}, Jennifer Karr\altaffilmark{1} and P. G. Martin}
\affil{Canadian Institute for Theoretical Astrophysics, University of
Toronto, Toronto, ON, Canada~M5S~3H8 \\ Electronic mail: ballanty, ferland, karr and pgmartin@cita.utoronto.ca}

\altaffiltext{1} {Also at: Department of Astronomy, University of Toronto, Toronto, ON, Canada~M5S~3H8} 
\altaffiltext{2}{Also at: Department of Physics, University of Kentucky, Lexington, KY~405506-0055 USA}

\begin{abstract}

It has long been possible to determine He$^+$/H$^+$ ratios from radio or infrared
recombination lines, making it possible to measure ionic abundance ratios
for
highly obscured emission-line objects. It is not possible to determine the
total
helium abundance from such data alone, however, since the ionization
correction
factor (ICF), the correction for unobservable stages of ionization of He or
H, must
also be determined.  Optical forbidden lines are usually used for this,
limiting
studies to relatively unobscured objects.

This paper outlines a way to determine the ICF using only infrared data.  We
identify four line pairs, [NeIII] 36~$\micron$/[NeII] 12.8~$\micron$, [NeIII]~15.6$\micron$ /[NeII] 12.8~$\micron$,
[ArIII] 9~$\micron$/[ArII] 6.9~$\micron$, and [ArIII] 21~$\micron$/[ArII] 6.9~$\micron$, that are sensitive
to the He
ICF. This happens because the ions cover a wide range of ionization, the
line
pairs are not sensitive to electron temperature, they have similar critical
densities, and are formed within the He$^+$/H$^+$ region of the nebula. We
compute
a very wide range of photoionization models appropriate for galactic HII
regions. The models cover a wide range of densities, ionization parameters,
stellar temperatures, and use continua from four very different stellar
atmospheres.

The results show that each line pair has a critical intensity ratio above
which
the He ICF is always small.  Below these values the ICF depends very
strongly on
details of the models for three of the ratios, and so other information
would be
needed to determine the helium abundance.  The [Ar III] 9~$\micron$/[ArII] 6.9~$\micron$ ratio
can
indicate the ICF directly due to the near exact match in the critical
densities of the
two lines.  Finally, continua predicted by the latest generation of stellar
atmospheres are sufficiently hard that they routinely produce significantly
negative ICFs.
\end{abstract}


\section{INTRODUCTION}

Helium abundances are fundamental tests of galactic nucleosynthesis (Pagel
1997, Shaver et al.\ 1983), the relationship between helium and metal
abundance
(dY/dZ), and its extrapolation to the primordial He/H ratio (Torres-Peimbert
et
al. 1989; Skillman et al.\ 1998).  The latter is a basic test of Big Bang
nucleosynthesis (Olive et al.\ 1997) and so has a cosmological imperative.
Great
precision is needed for cosmological tests since the range of He/H produced
in
different models of the Big Bang is not large (Olive \& Steigman 1995).  High
accuracy is possible since ratios of recombination lines are relatively
straightforward to convert into ionic abundance ratios (Peimbert 1975;
Benjamin
et al.\ 1999).

Several complications enter when great precision is needed. Collisional
excitation of optical helium lines can be important in denser objects
(Kingdon \&
Ferland 1996; Benjamin et al.\ 1999).  Accurate quantal calculations of the
relevant
rates now exist, and this can be taken into account if the density can be
determined.  Line transport effects can be significant for higher n radio
lines
(Goldberg 1966; Brocklehurst 1970).  The last, and most serious,
complication is
the ``Ionization Correction Factor'' (ICF), the correction for the fact that
(unobservable) atomic helium can be present in regions of a nebula where
hydrogen is ionized (Osterbrock 1989, Peimbert 1975). Again, other
spectroscopic
evidence, especially optical forbidden lines (Mathis 1982), can determine
the ICF.

Unfortunately it is not possible to convert the ionic He$^+$/H$^+$ ratio into a
total
He/H abundance without knowing the ICF, and so far only optical lines have
been used for this.  Radio and IR emission lines make it possible to map
He$^+$/H$^+$
across the Galaxy, or within heavily obscured environments such as starburst
galaxies (Shaver et al.\ 1983; Peimbert et al.\ 1988; Peimbert et al.\ 1992;
Simpson et al.\ 1995; Afflerbach et al.\ 1996; 1997; Rubin et al.\ 1998). It will soon be
possible to
routinely obtain high quality mid IR spectra of heavily obscured objects.
Methods of determining the He ICF from IR data alone, when other details of
the
source (density, geometry, stellar properties) are unknown, are needed. The
following investigates several line ratios that help make this possible

\section{CALCULATIONS}
\subsection{The Helium ICF}

Relative intensities of HeI to HI emission lines are proportional to the
ratio He$^+$/H$^+$, i.e.,
\begin{equation}
{I(HeI) \over I(HI)} = f(n_e,T_e){He^+ \over H^+}
\end{equation}
(Osterbrock 1989), where $f(n_e,T_e)$ incorporates the microphysics of the line
formation process (most recently summarized by Benjamin et al.\ 1999).   The
helium ionization correction factor (ICF) accounts for the presence of
atomic
helium in regions where hydrogen is ionized, and can be defined such that
\begin{equation}
{He \over H} = {He^+ \over H^+}(1+ \mathrm{ICF}).
\end{equation}
The ionic pair on the right hand side of Equation~1 can be measured with great
precision; the problem is to find a method to predict the ICF.  The ICF is
significant when the stellar continuum is too soft to maintain the helium
ionization across the hydrogen Str\"{o}mgren sphere. We expect the ICF to be
positive in most cases, since the He$^+$ volume is generally smaller than the
H$^+$ volume due to the higher ionization potential of He. It can be negative for
a very
hard continuum source since the photoionization cross section of atomic
helium
is larger than hydrogen at high energies (Shields 1974).
Photoionization models can predict the ionic fractions $A^{+}/A$ for any set of
parameters and geometry.  Defining a volume mean ionization fraction as
$<A^{+}/A>$, the helium ICF is given by
\begin{equation}
\mathrm{ICF}=\left< {H^+ \over H} \right> / \left< {He^+ \over He} \right> -1.
\end{equation}
Note that this allows for the possible presence of He$^{+2}$ in addition to He$^o$.
This will be the y-axis in most of the figures below.

\subsection{An Approach}

The aim of this paper is to identify line ratios that can indicate the ICF
using
IR data alone.  To be robust a method must not depend on details.  Our
approach
is to consider photoionization simulations over the broadest possible range
of
physical parameters, and then look for trends that are present in the
resulting
data set.  We took a similar approach to identify ways to determine the
bolometric luminosity using IR data alone (Bottorff et al.\ 1998).

We use the development version of the plasma simulation code Cloudy, last
described by Ferland et al.\ (1998).   The code used here is in excellent
agreement
with the methods, approximations, and results described in that review.

We concentrate on HII regions, since these best track the chemical evolution
of
the ISM.  We use Orion abundances (Rubin et al.\ 1991; Osterbrock et al.\
1992,
Baldwin et al.\ 1991).  A few of the abundances by number are He/H = 0.095,
C/H = 3.0(-4), N/H = 7.0 (-5), O/H=4.0 (-4), Ne/H=6.0 (-5) and Ar/H =3.0
(-6),
although all of the lightest thirty elements are included in our
calculation.

The assumed composition has little effect on our conclusions.  
As a test we computed a series of models in which the abundances were decreased to the very low metalicities
appropriate for low-mass galaxies.
The
main effect of the composition on the emission line spectrum is to change
the
equilibrium electron temperature (see Shields and Kennicutt 1995).  The
infrared
lines we use arise within the ground term of each ion, and so have very
small
excitation potentials and little temperature dependence (essentially $T^{-1/2}$
for each
line, so that temperature effects cancel in the ratio).  We compare only
ratios
involving different ionization stages of the same element.

We model a blister geometry, in which the HII region is an illuminated layer
on the face of a molecular cloud.  Specifically we assume constant density,
composition, and dust to gas ratio, across this layer. We use the Orion
grains and
the physics described by Baldwin et al.\ (1991).  These assumptions about the
grains introduce dependencies that are small compared to the dispersion of
results introduced by parameters we do vary, which are described next.

\subsection{Free Parameters}

The remaining parameters are the effective temperature and atmosphere of
the ionizing star, the gas density, and either the distance of the star to
the layer or
the ionization parameter $U$.  The ionization parameter is the dimensionless
ratio
of the density of hydrogen-ionizing photons at the illuminated face of the
cloud,
$\Phi(H)/c$, to the hydrogen density $n_H$,
\begin{equation}
U={ \Phi(H) \over n_H c}.
\end{equation}
The gas is more ionized, and the ICF tends to be smaller, for larger values
of $U$.

We expect that the helium ICF will be most sensitive to the stellar
temperature
and $U$, since these can selectively change the He$^+$/H$^+$ ratio.  The density has
little
affect on the ionization of helium relative to hydrogen when $U$ is held
constant,
although density does affect intensities of the IR lines when they are
collisionally
deexcited.

The ICF is most sharply dependent on the intensity of the helium-ionizing
stellar continuum ($h\nu > 1.8$~Ryd) relative to the hydrogen ionizing continuum
($h\nu > 1$~Ryd).  Figure~1 shows the four types of ionizing continua we use here,
all at
an effective temperature of 35,000~K and normalized to the same number of
hydrogen ionizing photons.  The first is a simple black body, which we take
as a
reference.  The Atlas LTE plane parallel atmosphere (Kurucz 1991) is
considerably softer than the blackbody at helium-ionizing energies.  The
Mihalas
(1972) NLTE plane parallel atmosphere is harder than the blackbody at the
highest energies, and depressed at intermediate energies, due to the opacity
of its
heavy pseudo-element.  Finally, the CoStar NLTE extended atmosphere
(Schaerer et al.\ 1996a, b) is the hardest of the set, with a flux equal to
or exceeding
the blackbody over much of the helium-ionizing continuum.   Of these four,
the
CoStar atmospheres are the only that include winds and full NLTE, and so may
be most realistic.

\subsection{Two Pairs of Line Rations}

The most likely candidates for robust ICF indicators are pairs of IR lines
of
different ionization stages of the same element. The ionization potentials
of both
stages must be greater than that of hydrogen to ensure that the lines form
in the
HII region (where the HeI and HI lines form) and not the background PDR (CI
recombination lines form here, Tielens \& Hollenbach 1985).  Ideally the
element
should be abundant (and not sharply depleted in the ISM or HII Regions) and
the lines in the pair should have similar critical densities (so that the
electron
density need not be well known).

The most promising possibilities are the two noble gasses Ne and Ar.  The
first
ion of each has an $np^{5}\phantom{i}^{2}\!P$ ground term and ionization potentials of 1.585~Ryd
and
1.158~Ryd for Ne and Ar respectively.  These produce the lines [Ne II]~$\lambda$~12.8~$\micron$
and [Ar II]~$\lambda$~6.9~$\micron$ with critical densities ($\log n$~cm$^{-3}$) of 5.8 and 5.6.  The second ion of each has an $np^{4}\phantom{i}^{3}\!P$ ground term, ionization potentials of 3.010~Ryd and 2.031~Ryd respectively, with each ion producing a pair of lines: [NeIII]~$\lambda\lambda$~36~$\micron$, 15.6~$\micron$ (critical densities, $\log n$, of 4.7 and 5.4) and [ArIII]~$\lambda\lambda$~9~$\micron$, 21~$\micron$ (critical densities 5.5 and 4.7).  We expect the [NeIII]~15.6~$\micron$/[NeII]~12.8~$\micron$ and [ArIII]~9~$\micron$/[Ar II]~6.9~$\micron$ ratios to be the best ICF indicators because of the similarities in their critical densities, with the Ar ratio the very best in this respect.
The Ne ions have larger ionization potentials than the Ar ions, so the Ne line ratios
may be a sharper indicator of the level of helium ionization.

\section{RESULTS}
\subsection{Results for Single Stellar Temperatures}

Figure~2 shows representative results for two stellar temperatures, 30,000~K
(the upper pair of panels) and 40,000~K (the lower pair).  CoStar
atmospheres
were used.  This temperature range is representative of objects with
non-trivial
He ICFs --- stars much hotter than 40,000~K easily sustain the helium
ionization,
while stars cooler than 30,000~K do not ionize helium at all.  The
ionization
parameter was varied between $U = 10^{-4}$ and 10$^{-1}$ to more than encompass the
range represented by spectra of observed HII regions.  The HII region
density
was varied between a density low enough to be in the asymptotic low-density
limit (10~cm$^{-3}$) and the high density of 10$^{6}$~cm$^{-3}$.  Most HII regions lie
below this
high density, and so this range should encompass nearly all objects.  The
range of
densities is broad enough to ensure that the full consequences of
collisional
deexcitation of the forbidden lines occur in many of the models.

The helium ICF is shown in the left pair of panels.  The dependence on the
stellar temperature is striking.  The ICF ranges between 2 and 20 for the
cool star,
but is small (and often negative) for the hot star.  A negative ICF occurs
when the
incident continuum is so hard that penetrating high-energy radiation
sustains the
ionization of He in regions where H is neutral (Shields 1974).
The ICF has a sharp dependence on $U$ and is nearly independent of the
density.  This is because the ionization parameter sets the overall
ionization of
the nebula.  The very weak density dependence is due to collisional effects
altering the fraction of helium triplet decays to ground - these produce
hydrogen-ionizing radiation and details can slightly change the resulting
geometry.

The predicted [NeIII]~36~$\micron$/[Ne II]~12.8~$\micron$ intensity ratios are shown in the
right
pair of panels. Only one of the four line pairs we consider is shown since
the
dependencies it exhibits are typical of the others. The line ratio has an
overall
dependence on $U$ that is similar to the ICF --- higher $U$ produces higher
ratios.
This is the underlying correlation we will exploit in the following.

Unfortunately the line ratio also has a significant density dependence for $n
> 10^4$~cm$^{-3}$, densities high enough to de-excite one but not both of the lines.
The
line ratio is density independent at densities substantially lower or higher
than
the critical densities of both lines.  It is clear from Figure~2 that gas
with densities $n \geq 10^4$~cm$^{-3}$ will introduce dispersion in the correlations we show next.

\subsection{Results for a Wide Range of Stellar Continua}

Grids similar to those shown in Figure~2 were computed for all four types of
stellar continua shown in Figure~1.  A broad range of temperatures was used
for
each stellar atmosphere --- 25,000~K to 50,000~K for the black body, 30,000~K to
50,000~K for CoStar, 30,000~K to 45,000~K for Atlas, and 30,000~K to 40,000~K for
Mihalas.  The ionization parameter and density were varied over the ranges
shown in Figure~2.  This resulted in a set of well over 10$^4$ independent
photoionization simulations. The results for [NeIII]/[NeII] are shown in
Figure~3 while Figure~4 shows [ArIII]/[ArII]. Results are presented with the
observable line ratio increasing to the right, and with the helium ICF, the quantity we
want to predict, as the dependent variable.

All four of the line ratios show a negative correlation: the helium ICF
generally increases as the line ratio decreases below a certain value.  The
line ratio decreases with decreasing levels of ionization, which in turn can be
due to either lower $U$ or stellar temperature (Figure~2).  This decreased ionization
correlates with larger amounts of atomic helium in the HII region, and so a
larger He ICF.

The range of densities we consider causes a large part of the scatter that
is present in the figures.  The line ratio can be small but the ionization of
the gas high if the double ionized species is collisionally de-excited.  This
explains why the [Ar III]~9~$\micron$/[Ar II]~6.9~$\micron$ ratio has the least scatter --- these lines happen to have nearly identical critical densities.  This line ratio could actually be used to deduce pathologically high values of the helium ICF.

For each line pair the ICF is small above some critical value of the ratio.
This suggests an observational approach.  One could easily identify those objects
for which this ratio is exceeded, and so, for whatever reason, have a small ICF.
For these the helium abundance is nearly equal to the measured He$^+$/H$^+$ ratio.
The critical line ratios are greater than unity.  In some cases the weaker line
might be unobservable, but the lower limit to the ratio is still of value to this
approach.

The ICF must be known to much better than 5\% accuracy to test Big Bang
nucleosynthesis. Figure~5 shows an expanded view of all four line ratios,
with a vastly expanded scale for the helium ICF. It is clear that a significant
amount of dispersion, at the several percent level, is still present.  Note also that
accurate determination of line ratios greater than ten may be difficult.  It is also
clear that the harder continua, the Mihalas and CoStar continua (both plotted as open
symbols), generally produce a \emph{negative} ICF.  If the CoStar continua are
representative of the spectra of windy stars, then negative ICFs could be a
major concern --- the helium abundance would be overestimated if this were not taken
into account.

\section{Conclusions}
\begin{itemize}

\item We have identified four IR line pairs that are sensitive to the helium
ionization correction factor.  These track the ionization of helium because
they are formed from adjacent stages of ionization of the same element. These
lines could be combined with radio or IR recombination lines to deduce a total
He/H abundance ratio.

\item For each line pair the helium ICF is smaller than several percent when the
ratio is above a certain critical intensity ratio.  Below this intensity
ratio the helium ICF may still be small in some cases (usually high density), but
other information, from other emission lines, would be needed to make progress.

\item Three of the line ratio/ICF correlations have large dispersion for a given
line ratio when the ICF is significant, due to differences in the critical
densities for each line.  The exception is the [ArIII]~9~$\micron$/[ArII]~6.9~$\micron$ 
ratio since these lines happen to have very similar critical densities.  For this pair a rough estimate of the ICF can be made from this line ratio alone.  In all four cases, the
estimated ICF could be made more accurate were other spectroscopic evidence available.

\item The prediction continua of the most recent generation of windy stellar
atmospheres (CoStar: Schaerer et al.\ 1996a, b), are sufficiently hard as to
produce a \emph{negative} helium ICF.  An observational analysis that did not allow
for this possibility would overestimate the helium abundance.
\end{itemize}

\acknowledgements

GJF thanks CITA for its hospitality during a sabbatical year, and
acknowledges support from the Natural Science and Engineering Research
Council of Canada through CITA.  DRB also acknowledges financial support
from NSERC.  Research in Nebular Astrophysics at the University of Kentucky
is supported by grants from NSF and NASA. We thank the referee for a careful
reading of the manuscript.

\clearpage

% Figure Captions - actual figure files are separate from MS

\figcaption{Examples of the four classes of ionizing continua used in these
calculations.  All had a temperature of 35,000~K and were normalized to the
same number of hydrogen ionizing photons.  The continua differ by several orders
of magnitude at the energies that sustain the He$^+$ ionization.}

\figcaption{The helium/hydrogen ionization correction factor and [NeIII]~36~$\micron$/[NeII]~12.8~$\micron$ intensity ratio for two stellar temperatures.  The quantities are plotted as a function of the hydrogen density and ionization parameter.  The continua were from CoStar atmosphere calculations for 30,000~K and 40,000~K. The major contour intervals are logarithmic in the two right-hand panels, and in the upper left panel, and linear in the lower left.}

\figcaption{The predicted line intensity ratios and helium/hydrogen ionization
correction factors for a very wide range of nebulae.  The ICF is near zero
for [NeIII]~36~$\micron$/[NeII]~12.8~$\micron$ intensity ratios greater than unity, and for [NeIII]~15.6~$\micron$/[NeII]~12.8~$\micron$ ratios greater than 10.}

\figcaption{The [ArIII]~9~$\micron$/[ArII]~6.9~$\micron$ and [ArIII]~21~$\micron$/[ArII]~6.9~$\micron$ intensity ratios, as in the previous figure.}

\figcaption{A vastly expanded view of the previous two figures.  A scatter in
the ICF of about 1 percent is present even at the highest line ratios.  Note the
frequently negative ICF for the Mihalas and CoStar spectra (plotted as open
symbols), which tend to be harder.}

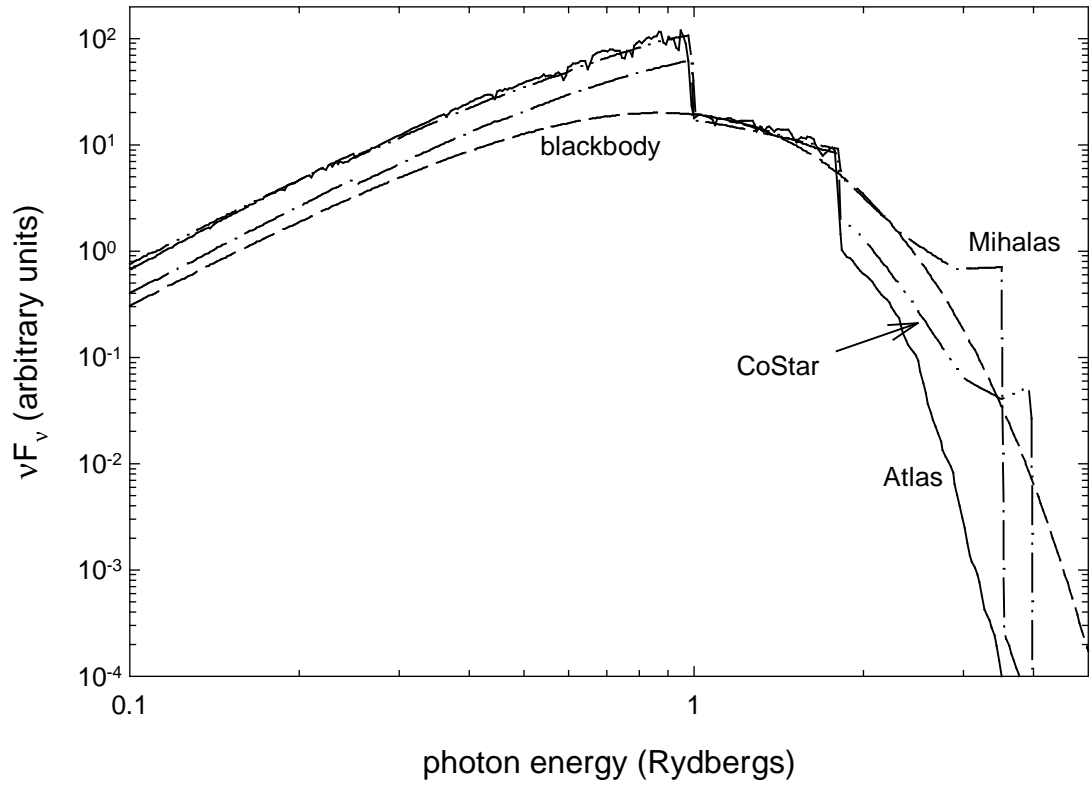
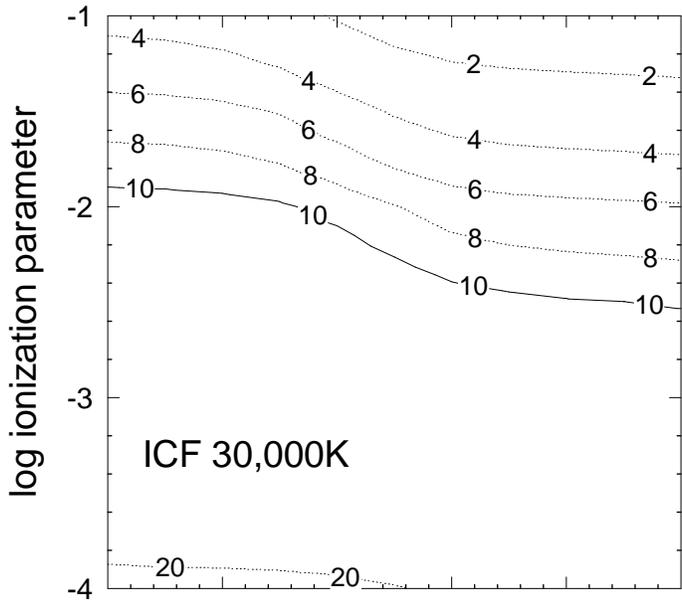
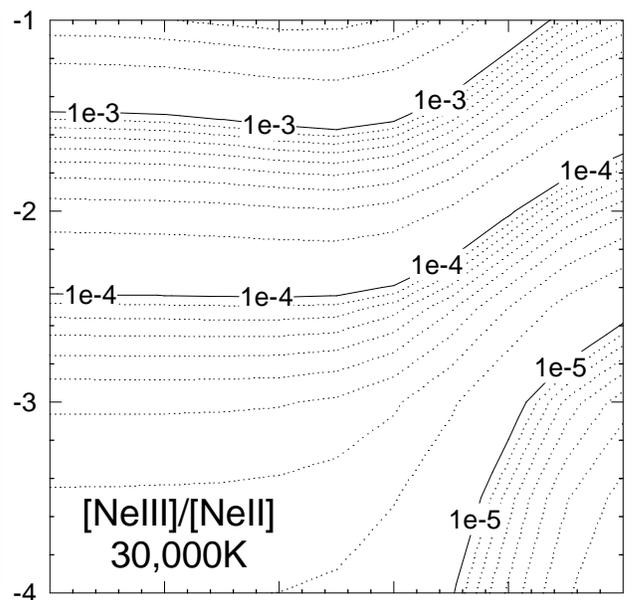
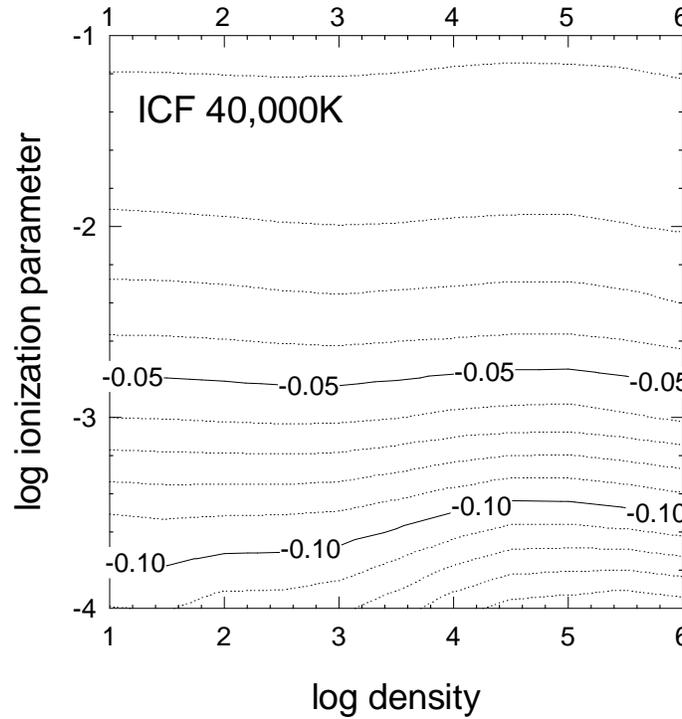
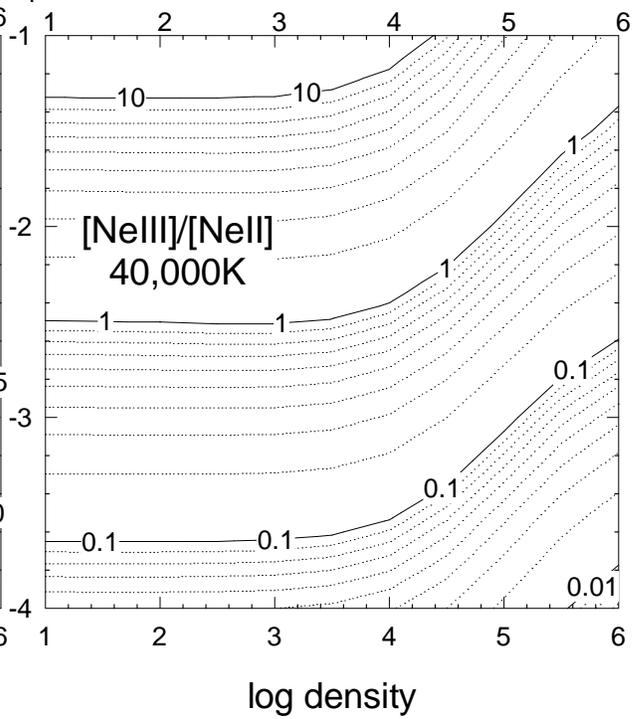
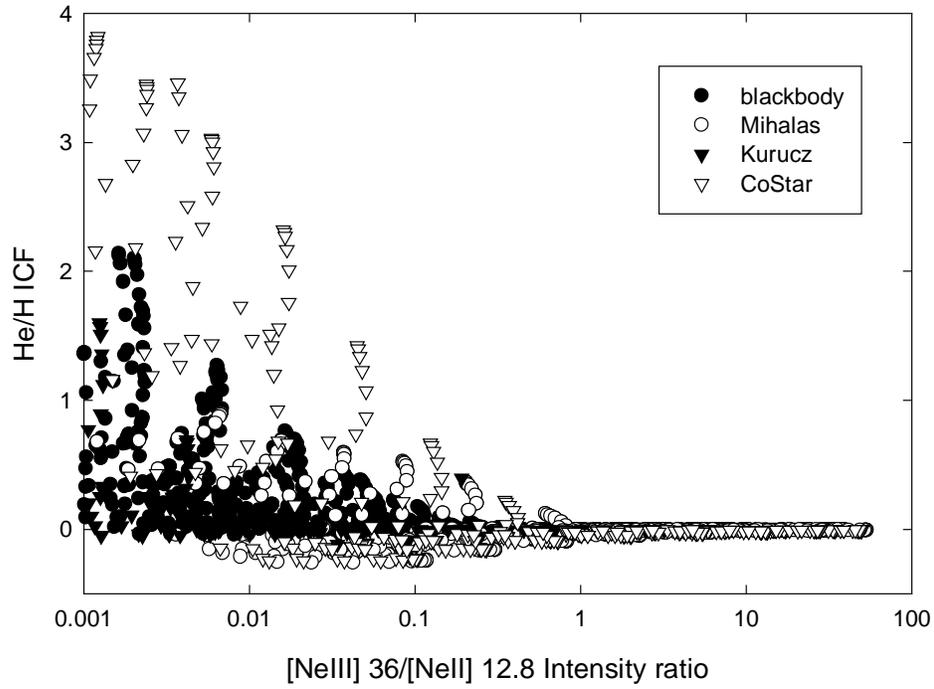
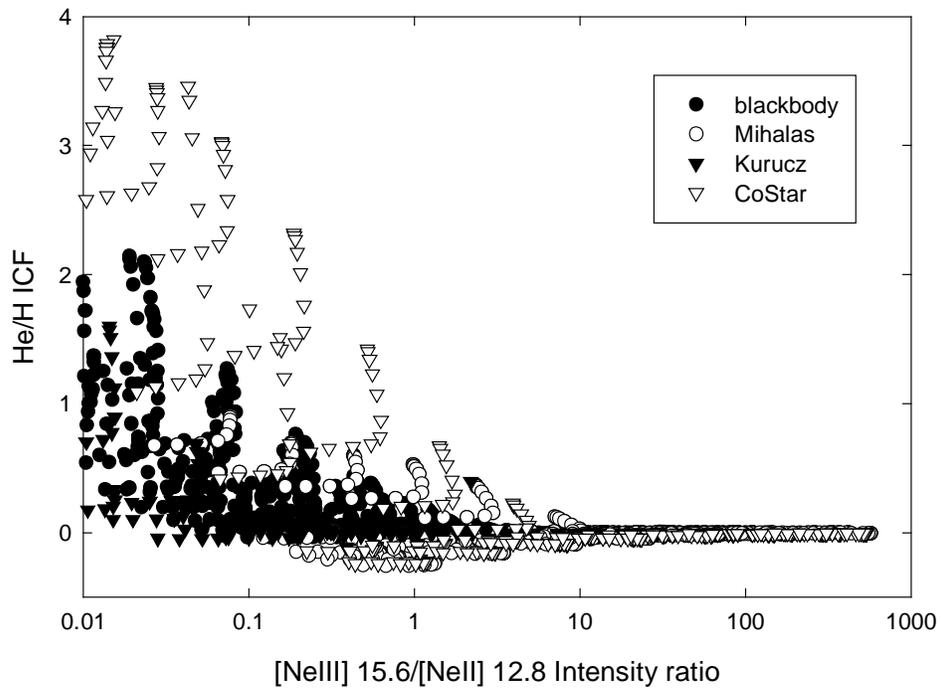
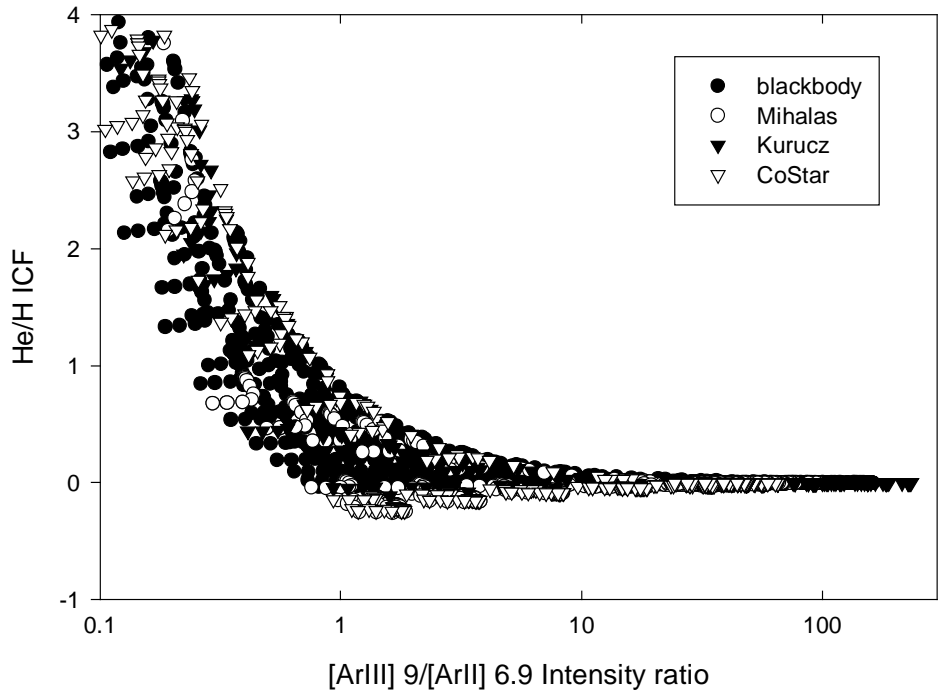
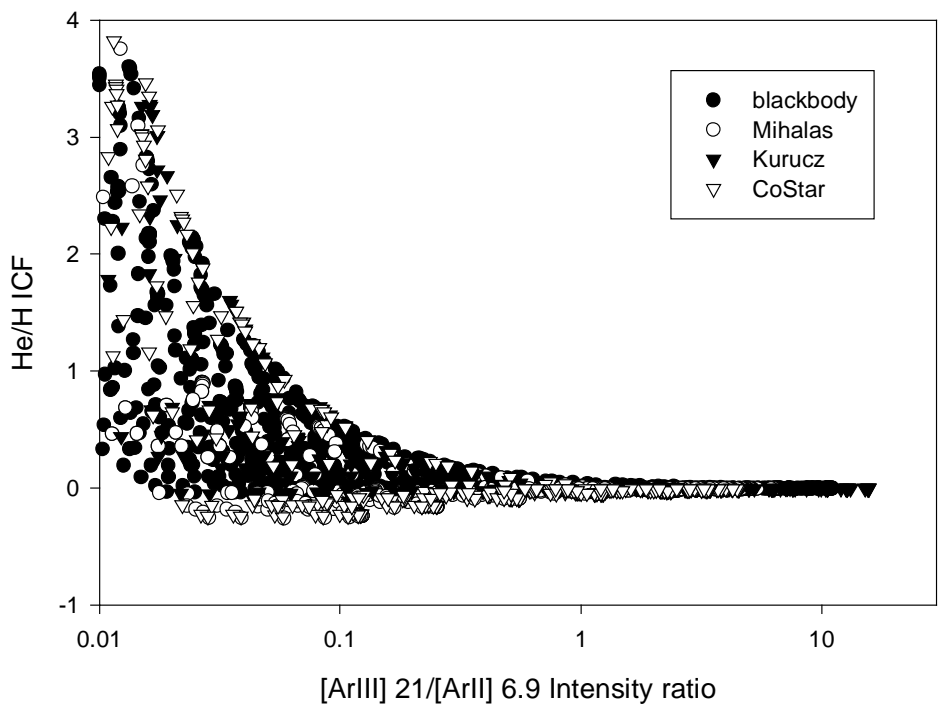
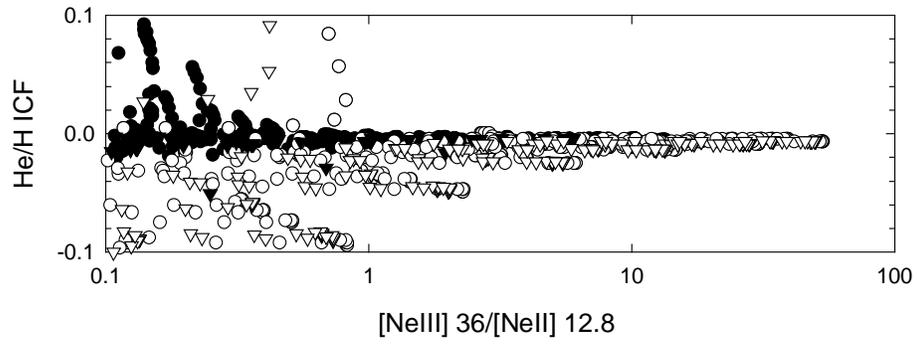
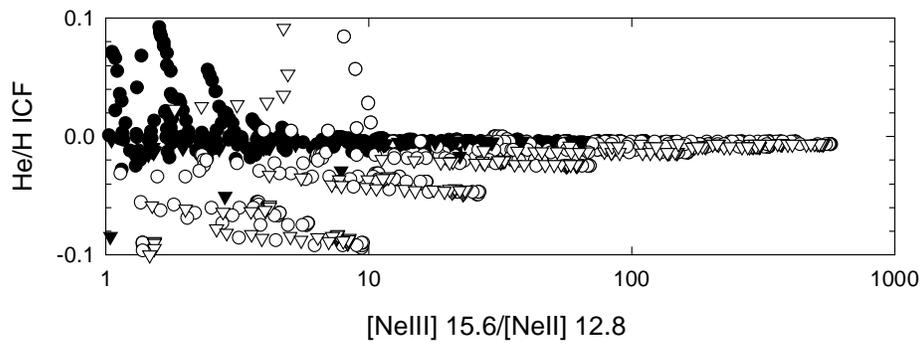
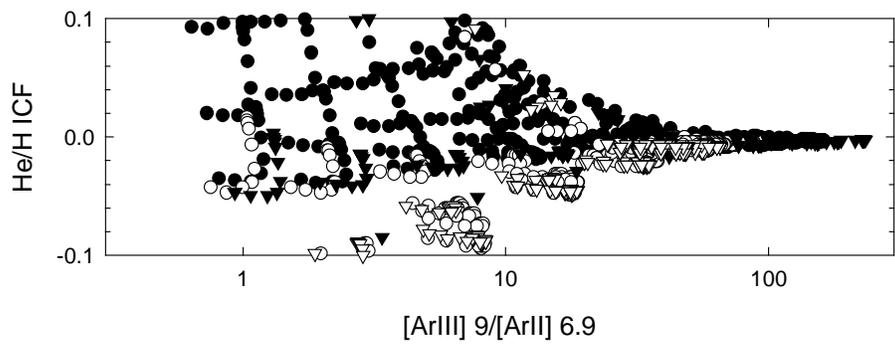
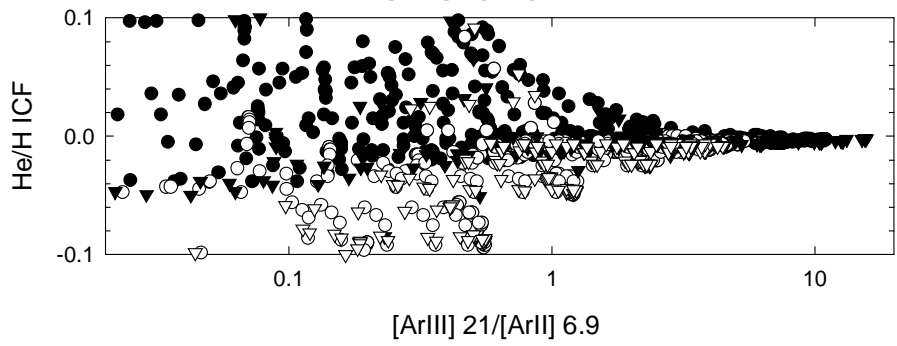
\end{document}